\documentclass[conference,a4paper]{IEEEtran}
\usepackage{xcolor}
\usepackage{balance}

\usepackage{cite}
\usepackage{amsmath,amssymb,amsfonts}
\usepackage{graphicx}
\usepackage{textcomp}
\usepackage{acronym}
\usepackage{xcolor}
\usepackage{tikz} 
\usepackage[utf8]{inputenc}
\usepackage{pgfplots} 
\usepackage{pgfgantt}
\usepackage{pdflscape}
\usepackage{changes}
\usepackage{comment}
\usepackage{subfigure}
\usepackage{mathtools,algpseudocode,algorithm,MnSymbol}
\usepackage{arydshln}
\usepackage{geometry}
\acrodef{MMSE}{Minimum Mean Squared Error}

\acrodef{MSE}{mean square error}

\acrodef{PSD}{power spectral density}

\acrodef{RMSE}{root mean squared error}
\acrodef{SLR}{statistical linear regression}

\acrodef{IPLF}{iterated posterior linearization filter}

\acrodef{DA}[DA]{data association}
\acrodef{ue}[UE]{user equipment}
\acrodef{bs}[BS]{base station}
\acrodef{va}[VA]{virtual anchor}
\acrodef{sp}[SP]{scattering  point}
\acrodef{fov}[FoV]{field-of-view}   
\acrodef{los}[LOS]{line-of-sight}
\acrodef{nlos}[NLOS]{non-line-of-sight}
\acrodef{PMBM}[PMBM]{Poisson  multi-Bernoulli  mixture}
\acrodef{PMB(M)}[PMB(M)]{Poisson  multi-Bernoulli  (mixture)}
\acrodef{PMB}[PMB]{Poisson  multi-Bernoulli}
\acrodef{rfs}[RFS]{random finite set}
\acrodef{PPP}[PPP]{Poisson point process}
\acrodef{MBM}[MBM]{multi-Bernoulli  mixture}
\acrodef{MB}[MB]{multi-Bernoulli}
\acrodef{ekf}[EKF]{extended Kalman filter}
\acrodef{ek}[EK]{extended Kalman}

\acrodef{PDF}[PDF]{probability density function}

\acrodef{ckf}[CKF]{cubature Kalman filter}
\acrodef{rbp}[RBP]{Rao-Blackwellized particle}
\acrodef{gospa}[GOSPA]{generalized optimal subpattern assignment}
\acrodef{slam}[SLAM]{simultaneous localization and mapping}
\acrodef{slat}[SLAT]{simultaneous localization and tracking}

\acrodef{TOA}[TOA]{time of arrival}
\acrodef{AOA}[AOA]{angles of arrival}
\acrodef{AOD}[AOD]{angles of departure}
\acrodef{CRB}[CRB]{Cram{\'e}r-Rao bound}
\acrodef{PCRB}[PCRB]{posterior Cram{\'e}r-Rao bound}
\acrodef{FIM}[FIM]{Fisher information matrix}
\acrodef{PIM}[PIM]{posterior information matrix}
\acrodef{PEB}[PEB]{position error bound}
\acrodef{LEB}[LEB]{landmark error bound}
\acrodef{HEB}[HEB]{heading error bound}
\acrodef{CEB}[CEB]{clock bias error bound}

\acrodef{MSE}[MSE]{mean squared error}
 
\usepackage{pgfplots}
  \pgfplotsset{compat=newest}
  \usetikzlibrary{plotmarks}
  \usetikzlibrary{arrows.meta}
  \usepgfplotslibrary{patchplots}
  \usepackage{grffile}
  \usepackage{amsmath}

\pgfplotsset{compat=newest} 
\pgfplotsset{plot coordinates/math parser=false} 

\DeclareMathOperator{\T}{\mathsf{T}}

\begin{document}

\bibliographystyle{IEEEtran}
\bstctlcite{IEEEexample:BSTcontrol}
%
\title{Doppler Exploitation in Bistatic\\ mmWave Radio SLAM}

\author{\IEEEauthorblockN{
Yu Ge\IEEEauthorrefmark{1}, Ossi Kaltiokallio\IEEEauthorrefmark{2},   
Hui Chen\IEEEauthorrefmark{1},
Fan Jiang\IEEEauthorrefmark{1}, Jukka Talvitie\IEEEauthorrefmark{2},\\
Mikko Valkama\IEEEauthorrefmark{2},
Lennart Svensson\IEEEauthorrefmark{1},      
Henk Wymeersch\IEEEauthorrefmark{1}      
}                                     
\IEEEauthorblockA{\IEEEauthorrefmark{1}
Department of Electrical Engineering, Chalmers University of Technology, Gothenburg, Sweden,\\ }
\IEEEauthorblockA{\IEEEauthorrefmark{2}
Unit of Electrical Engineering, Tampere University, Tampere, Finland,\\ }
\IEEEauthorblockA{ 
\{yuge,~hui.chen,~fan.jiang,~lennart.svensson,~henkw\}@chalmers.se,~\{ossi.kaltiokallio,~jukka.talvitie,~mikko.valkama\}@tuni.fi}}



\maketitle

\begin{abstract}
Networks in 5G and beyond utilize millimeter wave (mmWave) radio signals, large bandwidths, and large antenna arrays, which bring opportunities in jointly localizing the user equipment and mapping the propagation environment, termed as \ac{slam}. 
Existing approaches mainly rely on delays and angles, and ignore the Doppler, although it contains geometric information. In this paper, we study the benefits of exploiting Doppler in \ac{slam} through deriving the \acp{PCRB} and formulating the extended Kalman-Poisson multi-Bernoulli sequential filtering solution with Doppler as one of the involved measurements. Both theoretical \ac{PCRB} analysis and simulation results demonstrate the efficacy of utilizing Doppler. 
\end{abstract}

\vskip0.5\baselineskip

%

\section{Introduction}
MmWave communications in 5G and beyond are useful for \acf{slam} applications, due to geometrical propagation channels, large bandwidths, and large antenna arrays \cite{nurmi2017multi}. Signals sent from the \acf{bs} reach the \acf{ue} via the propagation channel, which is determined by the geometric relationships among the propagation environment, the \ac{ue}, and the \ac{bs}. Large bandwidths and antenna arrays result in high temporal and  spatial resolutions \cite{patwari2005,larsen2009}. Therefore, state-of-the-art channel estimators can provide accurate estimates for multipath components by using the received signals, in terms of groups of channel gain, \ac{TOA}, \ac{AOA}, \ac{AOD}, and Doppler, which contain  the necessary information for \ac{slam} \cite{wymeersch20175g}. Although Doppler contains geometric information, 
it is usually ignored. 

The related works can be divided into two areas: works that exploit Doppler for radio positioning or mapping and works in the area of radio SLAM. Doppler has been used in most radars for mapping and tracking \cite{patole2017automotive}, but limited works have been done in radio scenarios. Doppler is used to localize radio emitters in \cite{amar2008localization}, but the proposed method can only be used for narrow-band signals, while \cite{Han2016} shows that the Doppler shift can provide more direction information for localization, and \cite{kakkavas2019performance} shows that the mobility can significantly improve the \ac{nlos}-only scenario in MIMO mmWave system,  indicating the involvement of the Doppler shift brings gain in the localization accuracy. Doppler is used for tracking \acp{ue}' positions and velocities in a Wi-Fi-based system in  \cite{qian2017widar}. However, these methods do not solve the \ac{slam} problem. 
Several approaches have been proposed to address the mmWave radio \ac{slam} problem, including geometry-based methods \cite{wen20195g, yassin2018mosaic}, message passing-based methods \cite{Erik_BPSLAM_TWC2019,Rico_BPSLAM_JSTSP2019}, and \ac{rfs}-based methods \cite{kim20205g,ge20205GSLAM,ge2022computationally}. \ac{rfs}-based methods 
can  handle uncertainties, as well as inherently deal with challenges of the unknown number of landmarks, unknown \acp{DA}, misdetections, and clutter measurements in radio \ac{slam}. Within these \ac{rfs}-based methods, the probability hypothesis density (PHD) filter is used in \cite{kim20205g}, which does not have an explicit enumeration of \acp{DA}, and the \ac{PMBM} filter is used in \cite{ge20205GSLAM,ge2022computationally}, which enumerates all possible DAs explicitly, thus allowing for improved performance. However, the Doppler is not considered in these works. To our best knowledge, the inclusion of Doppler in radio SLAM has not yet been conducted in the existing literature.

In this paper, we harness the Doppler component in bistatic mmWave radio \ac{slam} and analyze how Doppler benefits the \ac{slam} filter. The main contributions of this paper are summarized as follows: (\emph{i}) we derive the \ac{PEB}, \ac{HEB}, \ac{CEB} of the \ac{ue} and the \acp{LEB} of landmarks by computing the \acf{PCRB} for the system with utilizing Doppler;  (\emph{ii}) we analyze the effect of the quality of the Doppler measurement on the bounds of the \ac{ue} and the average \ac{LEB} of landmarks; (\emph{iii}) we extend our previous work in \cite{ge2022computationally} by involving the Doppler as a dimension of the measurement in the  \ac{ek}-\ac{PMB} \ac{slam} filter, and validate the benefits of exploiting Doppler in the system through numerical experiments in the mmWave network context. 

\subsubsection*{Notations}
Scalars (e.g., $x$) are denoted in italic, vectors (e.g., $\boldsymbol{x}$) in bold lower-case letters  with $\left\Vert\boldsymbol{x}\right\Vert$ representing its L2-norm, matrices (e.g., $\boldsymbol{X}$) in bold capital letters, sets  (e.g., $\mathcal{X}$) in calligraphic with $\left|\mathcal{X}\right|$ representing its cardinality. The transpose is denoted by $(\cdot)^{\mathsf{T}}$, the Hermitian transpose is denoted by $(\cdot)^{\mathsf{H}}$, the union of mutually disjoint sets is denoted by $\uplus$, the Kronecker product is denoted by $\otimes$, the expectation is denoted by $\mathbb{E}[\cdot]$, $\mathcal{N}(\boldsymbol{u},\boldsymbol{\Sigma})$ denotes a multivariate Gaussian distribution with mean $\boldsymbol{u}$ and covariance $\boldsymbol{\Sigma}$, and $d_{\boldsymbol{x}}=\text{dim}(\boldsymbol{x})$.

\vspace{-1mm}
\section{System model}
\begin{figure}
\centering
    \includegraphics[width=0.98\linewidth]{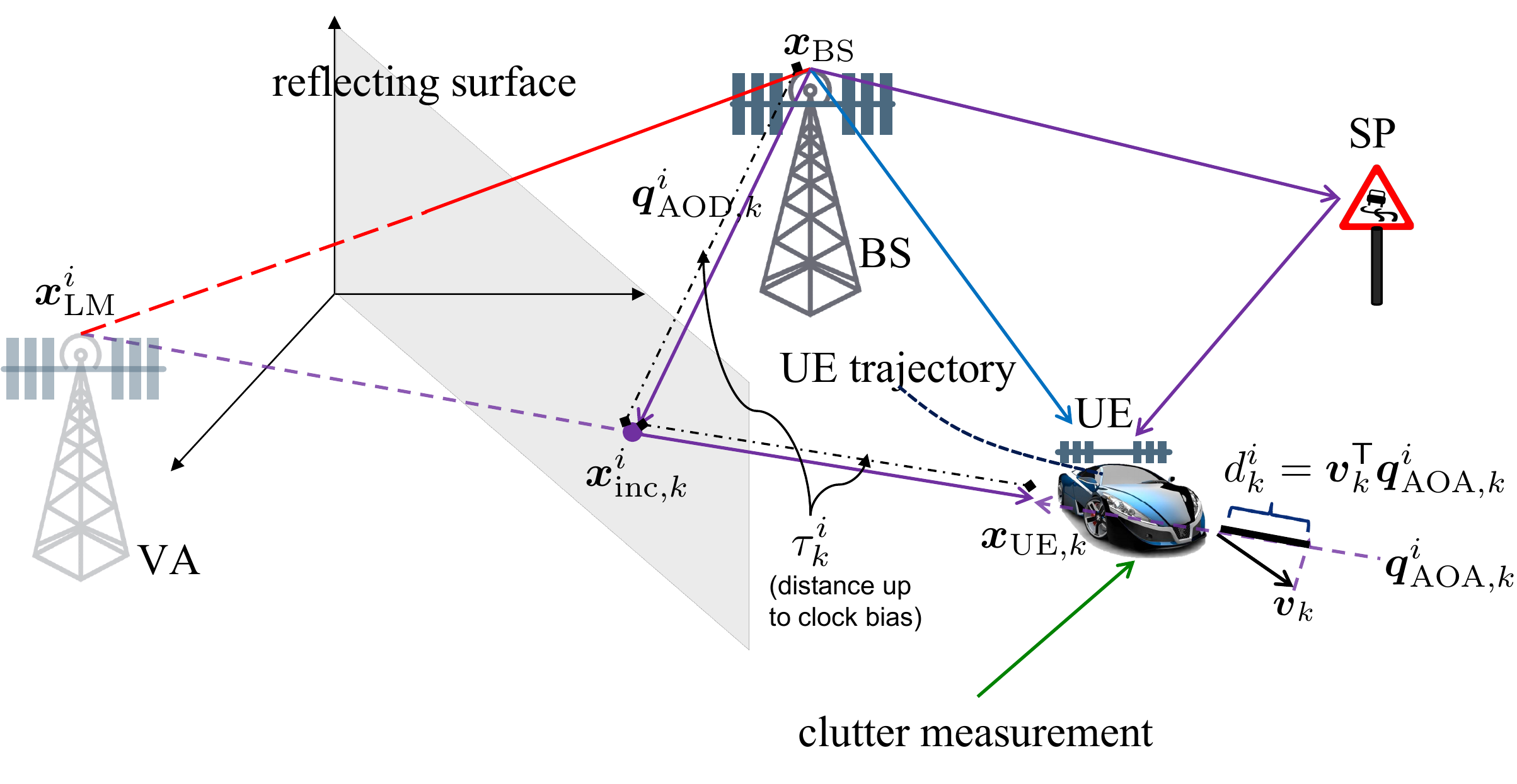}
\caption{A mmWave downlink scenario with the environment of a \ac{bs}, a \ac{ue}, a reflecting surface, and a small object. The reflecting surface can be modeled as a \acf{va}, which is the reflection of the \ac{bs} with respect to the surface. The small object is modeled as a \acf{sp}. The \ac{bs} sends signals to the \ac{ue} via \acf{los} path and/or \ac{nlos} paths, shown as the blue line and  purple lines, respectively. The channel parameters of each path depend on the underlying geometry.}
\label{fig:scenario}
\end{figure}

In this section, the \ac{ue} model, the environment model, the signal model, and the measurement model for a mmWave radio downlink scenario as shown in Fig.\,\ref{fig:scenario} are briefly introduced. 

\subsection{State Models}
In this paper, a single-user scenario is considered, where the \ac{ue} does a constant turn-rate movement around a known \ac{bs} on the $x-y$ plane.
We denote the dynamic state of the \ac{ue} at time step $k$ as $\boldsymbol{s}_{k}=[x_{k},y_{k},\varpi_{k},b_{k}]^{\mathsf{T}}$, containing the \ac{ue} position on $x$- and $y$-axis, heading, and clock bias, respectively. The \ac{ue} dynamics can be expressed as \cite{roth2014}
\begin{equation}\label{eq:dynamic_model}
     \underset{\boldsymbol{s}_{k}}{\underbrace{\begin{bmatrix} x_{k} \\ y_{k} \\ \varpi_{k} \\ b_{k} \end{bmatrix}}} = \underset{\boldsymbol{f}\left(\boldsymbol{s}_{k-1} \right)}{\underbrace{\begin{bmatrix} x_{k-1} + \tfrac{2v}{\omega}  \sin \! \left(\tfrac{\omega T}{2}\right)  \cos \! \left(\varpi_{k-1} + \tfrac{\omega T}{2}\right) \\ y_{k-1} + \tfrac{2v}{\omega}\sin \! \left(\tfrac{\omega T}{2}\right) \sin \! \left(\varpi_{k-1} + \tfrac{\omega T}{2}\right) \\ \varpi_{k-1} + \omega T \\ b_{k-1} \end{bmatrix}}} + \boldsymbol{\eta}_{k},
\end{equation}
where $v$, $\omega$, and $T$ are known control inputs, denoting the speed,  the turn-rate,  and the sampling interval, respectively, and  $\boldsymbol{\eta}_k \sim \mathcal{N}\left(\boldsymbol{0},\boldsymbol{Q}\right)$.

The considered scenario contains a single known \ac{bs}, and multiple unknown reflecting surfaces and small objects, which are modeled as \acp{va} and \acp{sp}, respectively. The \ac{bs} sends downlink signals to the \ac{ue}, which can reach the \ac{ue} directly, termed as the \ac{los} path, and/or can be reflected by the reflecting surfaces or scattered by the small objects and reach the  \ac{ue}, termed as \ac{nlos} paths  (see Fig.\,\ref{fig:scenario}). In this paper, we assume that there is at most one path associated to each landmark every time step. The landmark state  can be modeled as $\boldsymbol{x} = [\boldsymbol{x}^{\textsf{T}}_{\text{LM}},m]^{\textsf{T}}$,  with $\boldsymbol{x}_{\text{LM}} \in \mathbb{R}^{3}$ denoting the landmark location, and $m\in\{\text{BS},\text{VA},\text{SP}\}$ denoting the landmark type. Therefore, the map of the environment can be modeled by a set of landmark $\mathcal{X}=\{\boldsymbol{x}^{1},\dots, \boldsymbol{x}^{I}\}$, with ${I}$ representing the total number of landmarks.

\subsection{Signal Model}
\label{Signal model}
The \ac{bs} sends downlink signals to the \ac{ue} at every time step with a period of $T$ seconds. These signals can reach the \ac{ue} via \ac{los} path and/or \ac{nlos} paths, and the received signal for OFDM symbol $n$, at subcarrier $\kappa$ and time step $k$ can be expressed as~\cite{heath2016overview} 
\begin{align}
    \boldsymbol{y}_{\kappa,n,k} =\boldsymbol{W}_{n,k}^{\mathsf{H}}&\sum _{i=1}^{I_{k}}g_{k}^{i}\boldsymbol{a}_{\text{R}}(\boldsymbol{\theta}_{k}^{i})\boldsymbol{a}_{\text{T}}^{\mathsf{H}}(\boldsymbol{\phi}_{k}^{i}) e^{-\jmath 2\pi \kappa \Delta f \tau_{k}^{i}}\nonumber\\&\times e^{\jmath 2\pi n T_{\text{mea}}  d_{k}^{i}/\varsigma} \boldsymbol{p}_{n,k} + \boldsymbol{W}_{n,k}^{\mathsf{H}}\boldsymbol{n}_{\kappa,n,k}, \label{eq:FreqObservation} 
\end{align}
where $\boldsymbol{y}_{\kappa,n,k}$ denotes the received signal, $\boldsymbol{p}_{n,k}$ denotes the pre-coded pilot signal,  $\boldsymbol{n}_{\kappa,n,k}$ denotes a white Gaussian noise, $\boldsymbol{W}_{n,k}$ denotes a combining matrix, $I_{k}$ denotes the number of all visible landmarks at time $k$, $\boldsymbol{a}_{\text{R}}(\cdot)$ and $\boldsymbol{a}_{\text{T}}(\cdot)$ denote the steering vectors of the receiver and transmitter antenna arrays, respectively, $\Delta f$ denotes the subcarrier spacing, $T_{\text{mea}}$ denotes transmission duration of each symbol, and $\varsigma$ denotes the wavelength. 
Each path $i$ can be described by a complex gain $g_{k}^{i}$, a \ac{TOA} $\tau_{k}^{i}$, an \ac{AOA} pair $\boldsymbol{\theta}_{k}^{i}$ in azimuth and elevation,  an \ac{AOD} pair $\boldsymbol{\phi}_{k}^{i}$ in azimuth and elevation, and a Doppler $d_{k}^{i}$ which we express in [m/s]. These parameters can be  estimated by a parametric
channel estimation algorithm from $\boldsymbol{y}_{\kappa,n,k}$, for example, by \cite{richter2005estimation,alkhateeb2014channel,jiang2021high}, which is out of the scope of this paper, and the estimation results are utilized directly. Adding Doppler can make paths more resolvable, as there may be some paths not resolvable in delay or angle domains, but resolvable in the Doppler domain. 

\subsection{Measurement Model}
At time step $k$, the channel estimator provides a set of measurements $\mathcal{Z}_{k}=\{\boldsymbol{z}_{k}^{1},\dots, \boldsymbol{z}_{k}^{\hat{{I}}_{k}} \}$, where usually $\hat{{I}}_{k} \neq {{I}}_{k}$, as some visible landmarks may be misdetected and there could be some clutter measurements. Please note that the source of each element in $\mathcal{Z}_{k}$ is unknown.
If the measurement noise is zero-mean Gaussian, the measurement originating from landmark $\boldsymbol{x}^{i}$ at time step $k$ can be described as follows
\begin{align}
    \boldsymbol{z}^{i}_{k}=\boldsymbol{h}(\boldsymbol{x}^{i},\boldsymbol{s}_{k})+\boldsymbol{\epsilon}_k^i,\label{pos_to_channelestimation}
\end{align}
where  
$\boldsymbol{h}(\boldsymbol{x}^{i},\boldsymbol{s}_{k})=[\tau_{k}^{i},(\boldsymbol{\theta}_{k}^{i})^{\mathsf{T}},(\boldsymbol{\phi}_{k}^{i})^{\mathsf{T}}, d_{k}^{i}]^{\mathsf{T}}$ represents the nonlinear function that transforms the geometric information to the channel parameters, and $\boldsymbol{\epsilon}_k \sim \mathcal{N}\left(\boldsymbol{0},\boldsymbol{R}_{k}^{i}\right)$.

These channel parameters depend on the geometric relationships among the \ac{bs}, the \ac{ue} and the landmarks. Specifically, \ac{TOA} $\tau_{k}^{i}$ can be defined as
\begin{align}
    \tau_{k}^{i} = \begin{cases} \Vert\boldsymbol{x}_{\text{BS}}-\boldsymbol{x}_{\text{UE},k}\Vert/c + b_{k} & m^{i}=\text{BS} \\  (\Vert\boldsymbol{x}_{\text{inc},k}^{i}-\boldsymbol{x}_{\text{UE},k}\Vert + \Vert\boldsymbol{x}_{\text{inc},k}^{i}-\boldsymbol{x}_{\text{BS}}\Vert)/c +b_{k} & m^{i}\neq\text{BS},\end{cases} \label{delay}
\end{align}
where $\boldsymbol{x}_{\text{UE},k}= [x_{k},y_{k},0]^{\textsf{T}}$ denotes the 3D position of the \ac{ue}, $\boldsymbol{x}_{\text{inc},k}^{i}$ is the incidence point of the $i$-th path on the corresponding landmark at time step $k$, which can be determined by $\boldsymbol{x}_{\text{LM}}^{i}$ and $\boldsymbol{x}_{\text{UE},k}$, and $c$ is the speed of light. As $\boldsymbol{\theta}_{k}^{i}$ is determined by the arrival direction of the signal, which can be calculated by 
\begin{align}
    \boldsymbol{q}_{\text{AOA},k}^{i} = \begin{cases} (\boldsymbol{x}_{\text{BS}}-\boldsymbol{x}_{\text{UE},k})/\Vert\boldsymbol{x}_{\text{BS}}-\boldsymbol{x}_{\text{UE},k}\Vert & m^i=\text{BS} \\ (\boldsymbol{x}_{\text{inc},k}^{i}-\boldsymbol{x}_{\text{UE},k})/\Vert\boldsymbol{x}_{\text{inc},k}^{i}-\boldsymbol{x}_{\text{UE},k}\Vert & m^i\neq\text{BS},\end{cases}
\end{align}
we can define $\boldsymbol{\theta}_{k}^{i}$ as
\begin{align}
    &\theta_{\text{az},k}^{i} = \mathrm{arctan2}([\boldsymbol{q}_{\text{AOA},k}^{i}]_{2},[\boldsymbol{q}_{\text{AOA},k}^{i}]_{1})-\varpi_{k},\label{AOAAZ}\\
    &\theta_{\text{el},k}^{i} = \arcsin([\boldsymbol{q}_{\text{AOA},k}^{i}]_{3},\Vert\boldsymbol{q}_{\text{AOA},k}^{i}\Vert).\label{AOAEL}
\end{align}
Similarly, the departure direction can be calculated by  \begin{align}
    \boldsymbol{q}_{\text{AOD},k}^{i} = \begin{cases} (\boldsymbol{x}_{\text{UE},k}-\boldsymbol{x}_{\text{BS}})/\Vert(\boldsymbol{x}_{\text{UE},k}-\boldsymbol{x}_{\text{BS}})\Vert & m^i=\text{BS} \\ (\boldsymbol{x}_{\text{inc},k}^{i}-\boldsymbol{x}_{\text{BS}})/\Vert\boldsymbol{x}_{\text{inc},k}^{i}-\boldsymbol{x}_{\text{BS}}\Vert & m^i\neq\text{BS},\end{cases}
\end{align}
and $\boldsymbol{\phi}_{k}^{i}$ can be defined as
\begin{align}
    &\phi_{\text{az},k}^{i} = \mathrm{arctan2}([\boldsymbol{q}_{\text{AOD},k}^{i}]_{2},[\boldsymbol{q}_{\text{AOD},k}^{i}]_{1}),\label{AODAZ}\\
    &\phi_{\text{el},k}^{i} = \arcsin([\boldsymbol{q}_{\text{AOD},k}^{i}]_{3},\Vert\boldsymbol{q}_{\text{AOD},k}^{i}\Vert).\label{AODEL}
\end{align}
The Doppler $d_{k}^{i}$ can be calculated by projecting the \ac{ue} velocity on the direction of $\boldsymbol{q}_{\text{AOA},k}^{i}$, as displayed in Fig.\,\ref{fig:scenario}.  It can be computed by 
\begin{align}
    &d_{k}^{i} = \boldsymbol{v}_{k}^{\textsf{T}}\boldsymbol{q}_{\text{AOA},k}^{i},\label{Doppler}
\end{align}
where $\boldsymbol{v}_{k}$ denotes the 3D velocity of the \ac{ue} at time $k$, which can be defined as 
\begin{align}
    &\boldsymbol{v}_{k} = [v \cos \varpi_{k}, v\sin \varpi_{k}, 0]^{\mathsf{T}},\label{velocity}
\end{align}
since we assume the \ac{ue} only moves on the $x-y$ plane. Therefore, $d_{k}^{i}$ is positive, when the \ac{ue} approaches the $i$-th landmark, and it is negative, when the \ac{ue} moves away from the $i$-th landmark. 


\vspace{-1mm}

\section{Proposed Doppler-assisted SLAM}
In this section,  the \ac{PCRB} is briefly introduced,  the \ac{PEB}, \ac{HEB}, \ac{CEB} and \ac{LEB} are derived, and the contribution of Doppler to the \ac{slam} system is analyzed.


\subsection{\ac{PCRB}} \label{sec:performance_bounds}
The \ac{PCRB} is the lower bound that is analogous to the \ac{CRB} but takes a Bayesian perspective and assumes that there is a prior on the parameters, indicating that the \ac{MSE} of an estimator should always be larger than the inverse of the \ac{PIM} \cite[Ch. 4.2]{van2004detection}. Suppose $\boldsymbol{\alpha}$ denotes a vector of the measurements, $\boldsymbol{\beta}$ denotes an $r$-dimensional estimated random parameter, and $\boldsymbol{g}(\boldsymbol{\alpha})$ denotes an estimate of $\boldsymbol{\beta}$, which is a function of $\boldsymbol{\alpha}$. The \ac{PCRB} on the estimation error has the form
\begin{equation}\label{eq:cramer_rao_bound}
\mathbb{E}[ \left(\boldsymbol{g}(\boldsymbol{\alpha}) - \boldsymbol{\beta}\right)\left(\boldsymbol{g}(\boldsymbol{\alpha}) - \boldsymbol{\beta}\right)^{\mathsf{T}} ] \geq \boldsymbol{J}^{-1},
\end{equation}
where $\boldsymbol{J}$ is the $r\times r$ \ac{PIM} with elements
\begin{equation}\label{eq:cramer_rao_bound_elements}
[\boldsymbol{J}]_{\mu,\nu}=\mathbb{E}\lbrack-\frac{\partial^{2}\log f(\boldsymbol{\alpha},\boldsymbol{\beta})}{\partial {\beta}_{\mu} \partial {\beta}_{\nu}}\rbrack \qquad \mu,\nu=1,\cdots,r,
\end{equation}
with $f(\boldsymbol{\alpha},\boldsymbol{\beta})$ denoting the joint density. The inequality in \eqref{eq:cramer_rao_bound} indicates that the \ac{MSE} of an estimator is larger than $\boldsymbol{J}^{-1}$ in the positive semidefinite sense. The \ac{PIM} is a counterpart to the \ac{FIM} for the \ac{PCRB}, and can be decomposed into two  parts  $ \boldsymbol{J} = \boldsymbol{J}_{\text{data}} + \boldsymbol{J}_{\text{prior}}$, 
where $\boldsymbol{J}_{\text{data}}$  is the standard \ac{FIM}, which contains information obtained from the measurements, and $\boldsymbol{J}_{\text{prior}}$ is the priori information matrix, which contains prior information. The elements are given by (for $\mu,\nu=1,\cdots,r$) \vspace{-3mm}
\begin{align}
    \label{eq:cramer_rao_bound_elements_part1}
[\boldsymbol{J}_{\text{data}}]_{\mu,\nu}&=\mathbb{E}\lbrack-\frac{\partial^{2}\log f(\boldsymbol{\alpha}|\boldsymbol{\beta})}{\partial {\beta}_{\mu} \partial {\beta}_{\nu}}\rbrack, \\
\label{eq:cramer_rao_bound_elements_part2}
[\boldsymbol{J}_{\text{prior}}]_{\mu,\nu}&=\mathbb{E}\lbrack-\frac{\partial^{2}\log f(\boldsymbol{\beta})}{\partial {\beta}_{\mu} \partial {\beta}_{\nu}}\rbrack. 
\end{align}

\subsection{Performance Bounds} 

To compute the error bounds for the considered problem, we need to construct a  complete state of the system, denoted as $\breve{\boldsymbol{s}}_k = [(\boldsymbol{s}_k)^{\mathsf{T}}, \; (\boldsymbol{x}^{1}_{\text{LM},k})^{\mathsf{T}}, \; \ldots, \; (\boldsymbol{x}^{I}_{\text{LM},k})^{\mathsf{T}}]^{\mathsf{T}}$, the transition function of the complete state, denoted as $\breve{\boldsymbol{f}}(\breve{\boldsymbol{s}}_k) = [(\boldsymbol{f}(\boldsymbol{s}_{k})^{\mathsf{T}}, \; (\boldsymbol{x}^{1}_{\text{LM},k})^{\mathsf{T}}, \; \ldots, \; (\boldsymbol{x}^{I}_{\text{LM},k})^{\mathsf{T}}]^{\mathsf{T}}$, and the measurement function of the complete state given the ground-truth \ac{DA}, denoted as $\breve{\boldsymbol{h}}(\breve{\boldsymbol{s}}_k) = [(\bar{\boldsymbol{h}}(\boldsymbol{s}_{k},\boldsymbol{x}^{1}_{\text{LM},k})^{\mathsf{T}}, \; \ldots, \; (\bar{\boldsymbol{h}}(\boldsymbol{s}_{k},\boldsymbol{x}^{I}_{\text{LM},k})^{\mathsf{T}}]^{\mathsf{T}}$, with $\bar{\boldsymbol{h}}(\boldsymbol{s}_{k},\boldsymbol{x}^{i}_{\text{LM},k})=\boldsymbol{0}_{d_{\boldsymbol{z}} \times 1}$ when the landmark is not detected, and $\bar{\boldsymbol{h}}(\boldsymbol{s}_{k},\boldsymbol{x}^{i}_{\text{LM},k})$ is the same as $\boldsymbol{h}(\boldsymbol{s}_{k},\boldsymbol{x}^{i}_{\text{LM},k})$ when the landmark is detected. Then, if the ground-truth \ac{DA} is given, the \ac{PIM} can be recursively updated by \cite{tichavsky1998,EKPHD2021Ossi}
\begin{equation}\label{eq:fim_recursive}
\boldsymbol{J}_{k} = 
   \underset{\boldsymbol{J}_{\text{data},k}}{\underbrace{\breve{\boldsymbol{H}}_{k}^{\T} \breve{\boldsymbol{R}}_k^{-1}\breve{\boldsymbol{H}}_{k}}} +
   \underset{\boldsymbol{J}_{\text{prior},k}}{\underbrace{\left(\breve{\boldsymbol{Q}}_k + \breve{\boldsymbol{F}}_{k} \boldsymbol{J}_{k-1}^{-1} \breve{\boldsymbol{F}}_{k}^{\T} \right)^{-1}}}.
\end{equation} 
Here $\breve{\boldsymbol{H}}_{k}$ and $\breve{\boldsymbol{F}}_{k}$ are the Jacobian matrices of $\breve{\boldsymbol{h}}(\breve{\boldsymbol{s}}_k)$ and $\breve{\boldsymbol{f}}(\breve{\boldsymbol{s}}_k)$ with respect to $\breve{\boldsymbol{s}}_k$, evaluated at the true state, and $\breve{\boldsymbol{Q}}_k = \text{blkdiag}(\boldsymbol{Q},\boldsymbol{0}_{3I \times 3I})$ and $\breve{\boldsymbol{R}}_k = \text{blkdiag}(\boldsymbol{R}_k^1,\cdots,\boldsymbol{R}_k^I)$ denote the process and measurement noise covariances, respectively. Then, the \ac{PEB} can be computed as $\text{PEB}_k = \sqrt{[ \boldsymbol{J}_{k} ]_{1,1}^{-1} + [ \boldsymbol{J}_{k} ]_{2,2}^{-1}}$, the \ac{HEB} can be computed as $\text{HEB}_k = \sqrt{[ \boldsymbol{J}_{k} ]_{3,3}^{-1}}$, the \ac{CEB} can be computed as $\text{CEB}_k = \sqrt{[ \boldsymbol{J}_{k} ]_{4,4}^{-1}}$, and the \ac{LEB} of $i$-th landmark can be computed as $\text{LEB}_k^i = \sqrt{\sum_{\mu = 3i+2}^{3i + 4} [ \boldsymbol{J}_{k} ]_{\mu,\mu}^{-1}}$.

\subsection{Contribution of Doppler} \label{sec:contribution_Doppler}
From \eqref{eq:fim_recursive}, we can observe that the Doppler of the current measurements only contribute to $\boldsymbol{J}_{\text{data},k}$, as Doppler-related components are only involved in $\breve{\boldsymbol{H}}_{k}$ and $\breve{\boldsymbol{R}}_k$.  As $\breve{\boldsymbol{H}}_{k}$ is the Jacobian matrices of $\breve{\boldsymbol{h}}(\breve{\boldsymbol{s}}_k)$ with respect to $\breve{\boldsymbol{s}}_k$, it can be correspondingly decomposed into blocks
\begin{align}
    \breve{\boldsymbol{H}}_{k}  = \left[ \begin{array}{c : c c c c} \boldsymbol{A}_{k}^{1} & \boldsymbol{B}_{k}^{1}&\boldsymbol{0}_{6 \times 3}  &\cdots & \boldsymbol{0}_{6 \times 3}\\ \boldsymbol{A}_{k}^{2} & \boldsymbol{0}_{6 \times 3} & \boldsymbol{B}_{k}^{2} & \ddots & \boldsymbol{0}_{6 \times 3} \\ \vdots & \vdots& \ddots & \ddots & \vdots  \\ \boldsymbol{A}_{k}^{I} & \boldsymbol{0}_{6 \times 3}& \cdots& \boldsymbol{0}_{6 \times 3}& \boldsymbol{B}_{k}^{I}  \end{array}\right].
\end{align}
Here, $\boldsymbol{A}_{k}^{i}$ and $\boldsymbol{B}_{k}^{i}$ are the Jacobian matrices of $\bar{\boldsymbol{h}}(\boldsymbol{s}_{k},\boldsymbol{x}^{i}_{\text{LM}})$ with respect to $\boldsymbol{s}_{k}$ and $\boldsymbol{x}^{i}_{\text{LM}}$, respectively, which are $\boldsymbol{0}_{6\times4}$ and $\boldsymbol{0}_{6\times3}$ when the $i$-th landmark is not detected, and are
\begin{align}
    \boldsymbol{A}_{k}^{i} =\left[ \begin{array}{c}
    \tilde{\boldsymbol{A}}_{k}^{i}\\ \hdashline  (\tilde{\boldsymbol{a}}_{k}^{i})^{\mathsf{T}}
    \end{array}\right], \quad \boldsymbol{B}_{k}^{i} =\left[ \begin{array}{c}
    \tilde{\boldsymbol{B}}^{i}_{k}\\ \hdashline \vspace{-4mm} \\ (\tilde{\boldsymbol{b}}_{k}^{i})^{\mathsf{T}}
    \end{array}\right],
\end{align}
when the landmark is detected, where $ \tilde{\boldsymbol{A}}_{k}^{i}$ is the Jacobian of $[\tau_{k}^{i},(\boldsymbol{\theta}_{k}^{i})^{\mathsf{T}},(\boldsymbol{\phi}_{k}^{i})^{\mathsf{T}}]^{\mathsf{T}}$ with respect to $\boldsymbol{s}_{k}$, $ \tilde{\boldsymbol{B}}^{i}_{k}$ is the Jacobian of $[\tau_{k}^{i},(\boldsymbol{\theta}_{k}^{i})^{\mathsf{T}},(\boldsymbol{\phi}_{k}^{i})^{\mathsf{T}}]^{\mathsf{T}}$ with respect to $\boldsymbol{x}^{i}_{\text{LM}}$, and 
\begin{align}
    (\tilde{\boldsymbol{a}}_{k}^{i})^{\mathsf{T}} & =[
    \begin{array}{cccc}
         \frac{\partial d_{k}^{i}}{\partial x_{k}} & \frac{\partial d_{k}^{i}}{\partial y_{k}} &\frac{\partial d_{k}^{i}}{\partial \varpi_{k}} & \frac{\partial d_{k}^{i}}{\partial b_{k}}
    \end{array}
    ],\\
    (\tilde{\boldsymbol{b}}_{k}^{i})^{\mathsf{T}} &  =[
    \begin{array}{cccc}
        \frac{\partial d_{k}^{i}}{\partial x_{\text{LM}}^{i}} & \frac{\partial d_{k}^{i}}{\partial y_{\text{LM}}^{i}} &\frac{\partial d_{k}^{i}}{\partial z_{\text{LM}}^{i}}
     \end{array}
    ].
\end{align}
Furthermore, we assume $\boldsymbol{R}_{k}^{i}=\text{diag}(\tilde{\boldsymbol{R}}_{k}^{i},(\sigma_{\text{d},k}^{i})^{2})$. 
  According to \eqref{eq:fim_recursive} and the inverse of a block matrix, $\boldsymbol{J}_{\text{data},k}$ is only non-zero in the diagonal block related to the UE state
\begin{align}
&[\boldsymbol{J}_{\text{data},k}]_{1:4,1:4}= \sum_{i=1}^{I}(\boldsymbol{A}_{k}^{i})^{\mathsf{T}}(\boldsymbol{R}_{k}^{i})^{-1}\boldsymbol{A}_{k}^{i}\nonumber\\&= 
\sum_{i=1}^{I}\underbrace{(\tilde{\boldsymbol{A}}_{k}^{i})^{\mathsf{T}}(\tilde{\boldsymbol{R}}_{k}^{i})^{-1}\tilde{\boldsymbol{A}}_{k}^{i}}_{\text{non-Doppler related}}+\sum_{i=1}^{I}\underbrace{\tilde{\boldsymbol{a}}_{k}^{i}(\sigma_{\text{d},k}^{i})^{-2}(\tilde{\boldsymbol{a}}_{k}^{i})^{\mathsf{T}}}_{\text{Doppler related}},\label{eq:cramer_rao_bound_elements_data_decomposeA}
\end{align}
and the diagonal blocks related to each landmark
\begin{align}
&[\boldsymbol{J}_{\text{data},k}]_{3i+2:3i+4,3i+2:3i+4}= (\boldsymbol{B}_{k}^{i})^{\mathsf{T}}(\boldsymbol{R}_{k}^{i})^{-1}\boldsymbol{B}_{k}^{i}\nonumber\\&= \underbrace{
(\tilde{\boldsymbol{B}}_{k}^{i})^{\mathsf{T}}(\tilde{\boldsymbol{R}}_{k}^{i})^{-1}\tilde{\boldsymbol{B}}_{k}^{i}}_{\text{non-Doppler related}}+\underbrace{\tilde{\boldsymbol{b}}_{k}^{i}(\sigma_{\text{d},k}^{i})^{-2}(\tilde{\boldsymbol{b}}_{k}^{i})^{\mathsf{T}}}_{\text{Doppler related}}.\label{eq:cramer_rao_bound_elements_data_decomposeB}
\end{align}
Clearly, Doppler provides non-negative information to both UE and landmark states. Due to the block-diagonal structure of 
$\boldsymbol{J}_{\text{data},k}$, \eqref{eq:cramer_rao_bound_elements_data_decomposeA} and \eqref{eq:cramer_rao_bound_elements_data_decomposeB} can be interpreted as the equivalent FIMs of the UE and landmark states. 

To gain further insights, we expand
$\tilde{\boldsymbol{a}}^i_{k}$ and $\tilde{\boldsymbol{b}}^i_{k}$ as 
\begin{align}
    & \tilde{\boldsymbol{a}}_{k}^{i}=\big[-\frac{\left([\boldsymbol{v}_{k}]_{1:2}-d_{k}^{i}[\boldsymbol{q}_{\text{AOA},k}^{i}]_{1:2}\right)^{\mathsf{T}}}{\Vert\boldsymbol{x}_{\text{inc},k}^{i}-\boldsymbol{x}_{\text{UE},k}\Vert}, \boldsymbol{v}_{\perp,k}^{\textsf{T}}\boldsymbol{q}_{\text{AOA},k}^{i},0 \big]^{\mathsf{T}}, \\
    & \tilde{\boldsymbol{b}}_{k}^{i} =\frac{\partial \boldsymbol{x}_{\text{inc},k}^{i}}{\partial \boldsymbol{x}_{\text{LM}}^{i}}\frac{\boldsymbol{v}_{k}-d_{k}^{i}\boldsymbol{q}_{\text{AOA},k}^{i}}{\Vert\boldsymbol{x}_{\text{inc},k}^{i}-\boldsymbol{x}_{\text{UE},k}\Vert}, 
\end{align}
where $\boldsymbol{v}_{\perp,k}=[ -v\sin \varpi_{k}, v\cos \varpi_{k}, 0]^{\mathsf{T}}$ (i.e., $\boldsymbol{v}^\top_{\perp,k}\boldsymbol{v}_{k}=0$). Then,  the additive Doppler-related parts in \eqref{eq:cramer_rao_bound_elements_data_decomposeA} and \eqref{eq:cramer_rao_bound_elements_data_decomposeB} can be summarized as follows:
\begin{itemize}
\item \emph{\ac{ue} position:} as $\boldsymbol{v}_{k}-d_{k}^{i}\boldsymbol{q}_{\text{AOA},k}^{i}=\boldsymbol{v}_{k}-\boldsymbol{v}_{k}^{\textsf{T}}\boldsymbol{q}_{\text{AOA},k}^{i}\boldsymbol{q}_{\text{AOA},k}^{i}$, the  information brought by Doppler is along the direction of the velocity, when the velocity is orthogonal to a path's $\boldsymbol{q}_{\text{AOA},k}^{i}$ direction, i.e., when $\boldsymbol{v}_{k}^{\textsf{T}}\boldsymbol{q}_{\text{AOA},k}^{i}=0$. On the other hand, when the velocity is parallel to a path's $\boldsymbol{q}_{\text{AOA},k}^{i}$ direction, the Doppler of that path does not provide direct Fisher information.  A similar argument holds for landmark locations.


\item \emph{Heading $\varpi_{k}$:} the direct contribution is from the projection of $\boldsymbol{v}_{\perp,k}$ on the direction of $\boldsymbol{q}_{\text{AOA},k}^{i}$ of each path. Hence, if the \ac{ue} velocity is parallel to a path's $\boldsymbol{q}_{\text{AOA},k}^{i}$ direction, 
the direct Fisher information  is 0.

\item \emph{Clock bias $b_{k}$:} there is no direct Fisher information contribution. However, the direct contributions to the other dimensions still benefit estimate of the clock bias.  
\end{itemize}
Therefore, if the \ac{ue} moves alongside the same or the opposite direction of the  $\boldsymbol{q}_{\text{AOA},k}^{i}$, the corresponding group of measurement  does not  has any Doppler-related information. Moreover, a smaller $(\sigma_{\text{d},k}^{i})^{2}$ results in larger additional Doppler-related information.

\section{EK-PMB(M) SLAM filter}
The \ac{slam} framework in this paper follows the EK-PMB \ac{slam} filter proposed in \cite{ge2022computationally}. In this section, the basics of the PMB(M) density and the EK-PMB(M) SLAM filter are summarized in the following. 

\subsection{Basics of PMB(M) Density}
The map of the environment $\mathcal{X}$ is formulated as a \ac{PMBM} \ac{rfs}, which can be viewed as the union of two disjoint \acp{rfs}: $\mathcal{X}_{\mathrm{U}}$ for the set of undetected landmarks, which are the landmarks that have never been detected, and $\mathcal{X}_{\mathrm{D}}$ for the set of detected landmarks, which are the the landmarks that have been detected at least once before. Its density follows \cite{garcia2018poisson}
\begin{equation}
    f(\mathcal{X})=\sum_{\mathcal{X}_{\mathrm{U}}\biguplus\mathcal{X}_{\mathrm{D}}=\mathcal{X}}f_{\mathrm{P}}(\mathcal{X}_{\mathrm{U}})f_{\mathrm{MBM}}(\mathcal{X}_{\mathrm{D}}),\label{PMBM}
\end{equation}
where $\mathcal{X}_{\mathrm{U}}$ and  $\mathcal{X}_{\mathrm{D}}$ are modeled as a \ac{PPP} and a \ac{MBM}, respectively, with densities following
\begin{align}
    f_{\mathrm{P}}(\mathcal{X}_{\mathrm{U}})&=e^{-\int\lambda(\boldsymbol{x})\mathrm{d}\boldsymbol{x}}\prod_{\boldsymbol{x} \in \mathcal{X}_{\mathrm{U}} }\lambda(\boldsymbol{x}),\label{PPP} \\
     f_{\mathrm{MBM}}(\mathcal{X}_{\mathrm{D}})&\propto \sum_{j \in \mathbb{I}}w^{j}\sum_{\mathcal{X}^{1}\biguplus \dots \biguplus \mathcal{X}^{|\mathcal{X}_{\mathrm{D}}|}=\mathcal{X}_{\mathrm{D}}}\prod_{i=1}^{|\mathcal{X}_{\mathrm{D}}|}f^{j,i}_{\mathrm{B}}(\mathcal{X}^{i}).\label{MBM}
\end{align}
Here, $\lambda(\cdot)$ denotes the intensity function of the \ac{PPP} density, and  $\mathbb{I}$ is the index set of all global hypotheses with weights satisfying $\sum_{j\in\mathbb{I}}w^{j}=1,w^{j}\ge 0$, and the global hypotheses in \ac{slam} correspond to different \acp{DA} \cite{williams2015marginal}. Each individual component in \eqref{MBM} is termed as a Bernoulli process, and $f_{\mathrm{B}}^{j,i}(\cdot)$ denotes the Bernoulli density of the $i$-th landmark under the $j$-th global hypothesis, following
\begin{equation}
f^{j,i}_{\mathrm{B}}(\mathcal{X}^{i})=
\begin{cases}
1-r^{j,i} \quad& \mathcal{X}^{i}=\emptyset \\ r^{j,i}f^{j,i}(\boldsymbol{x}) \quad & \mathcal{X}^{i}=\{\boldsymbol{x}\} \\ 0 \quad & \mathrm{otherwise},
\end{cases}
\end{equation} 
where $f^{j,i}(\cdot)$ denotes the state density, and $r^{j,i} \in [0,1]$ denotes the probability that the corresponding landmark exists. 
We note that if there is only one mixture component in the \ac{MBM}, then \eqref{MBM} reduces to a \ac{MB}, and \eqref{PMBM} reduces to a \ac{PMB}.  


\subsection{ EK-PMB(M) SLAM Filter Recursion} \label{EK-PMB}
The EK-PMB(M) SLAM filter follows the Bayesian filtering recursion with \acp{rfs}, and it computes the joint posterior $f(\boldsymbol{s}_{k+1},\mathcal{X}|\mathcal{Z}_{1:k+1})$  by \cite{ge2022computationally}
\begin{align}
     f(\boldsymbol{s}_{k+1},\mathcal{X}|\mathcal{Z}_{1:k+1}&) \propto   \ell(\mathcal{Z}_{k+1}|\boldsymbol{s}_{k+1},\mathcal{X}) f(\mathcal{X}|\mathcal{Z}_{1:k}) \nonumber \\ &\times f(\boldsymbol{s}_{k+1}|\mathcal{Z}_{1:k}), \label{joint_posterior}
\end{align}
with $f(\boldsymbol{s}_{k+1}|\mathcal{Z}_{1:k})=\int f(\boldsymbol{s}_{k}|\mathcal{Z}_{1:k})f(\boldsymbol{s}_{k+1}|\boldsymbol{s}_{k}) \text{d} \boldsymbol{s}_{k}$ and $\ell(\mathcal{Z}_{k+1}|\boldsymbol{s}_{k+1},\mathcal{X})$ denoting the RFS likelihood function, given by  \cite[eqs.\,(5)--(6)]{garcia2018poisson}. Instead of tracking the joint density, the filter keeps track of the marginal  \ac{ue} $f(\boldsymbol{s}_{k}|\mathcal{Z}_{1:k})$ and map $f(\mathcal{X}|\mathcal{Z}_{1:k})$ posteriors by marginalizing out the map and the \ac{ue}  state from the joint posterior, respectively. The marginal posteriors are given by
\begin{align}
     f(\boldsymbol{s}_{k+1}|\mathcal{Z}_{1:k+1})  &= \int  f(\boldsymbol{s}_{k+1},\mathcal{X}|\mathcal{Z}_{1:k+1})\delta \mathcal{X},  \label{eq:UpMarObjMarg1}\\
    f(\mathcal{X}|\mathcal{Z}_{1:k+1}) &= \int f(\boldsymbol{s}_{k+1},\mathcal{X}|\mathcal{Z}_{1:k+1}) \mathrm{d}\boldsymbol{s}_{k+1},\label{eq:UpMarObjMarg2}
\end{align}
where $\int \psi(\mathcal{X})\delta \mathcal{X}$ refers to the set integral \cite[eq.~(4)]{williams2015marginal}.  

The EK-PMB(M) \ac{slam} filter proposed in \cite{ge2022computationally}  determines $\gamma\ge 1$ most likely \acp{DA} for each prior global hypothesis with corresponding weights, and the joint posterior of the \ac{ue} state and the map conditioned on each \ac{DA} is computed by utilizing an \ac{ekf}-like update step. Then the joint posterior in \eqref{joint_posterior} can be acquired by the weighted summation of the densities for \acp{DA}, followed by \eqref{eq:UpMarObjMarg1} and \eqref{eq:UpMarObjMarg2} to compute the \ac{ue} state and the map. To avoid the exponential increase of \acp{DA}, we can approximate the PMBM density to a PMB density at the end of each update step, which reduces the complexity significantly.

It is important to mention that the weights of $\gamma\ge 1$ most likely \acp{DA} for each prior global hypothesis are computed by using measurements \cite[eq.\,(29)]{ge2022computationally}. Since more information is provided with the involvement of Doppler, finding the correct \acp{DA} becomes more likely, 
as correct local hypothesis weights computed in \cite[eqs.\,(22)--(24)]{ge2022computationally} become more prominent.

\section{Results}

\subsection{Simulation Environment}
We consider a 5G downlink scenario with a single \ac{bs} located at $[0 \, \text{m},0 \, \text{m},40\, \text{m}]^{\mathsf{T}} $, 4 \acp{va} located at  $[\pm 200 \, \text{m},0 \, \text{m},40 \, \text{m}]^{\mathsf{T}}$, $[0 \, \text{m}, \pm 200 \, \text{m},40 \, \text{m}]^{\mathsf{T}}$, and 4 \acp{sp} located at  $[\pm 99 \, \text{m},0 \, \text{m},10 \, \text{m}]^{\mathsf{T}}$, $[0 \, \text{m}, \pm 99 \, \text{m},10 \, \text{m}]^{\mathsf{T}}$. The \ac{ue} does a counterclockwise  constant turn-rate movement around the \ac{bs} according to \eqref{eq:dynamic_model}, with $v = 22.22 \text{ m/s}$,  $\omega = \pi/10 \text{ rad/s}$, $T=0.5 \text{ s}$, and $\mathbf{Q}= \text{diag}(0.2^2 \text{ m}^2, \,0.2^2 \text{ m}^2, \,0.001^2 \text{ rad}^2, \,0.2^2 \text{ m}^2)$, and it takes $K = 40$ samples for the \ac{ue} to circle the road once. The measurement covariance matrix is set as  $\boldsymbol{R}_{k}^{i} = \text{blkdiag}(10^{-2} \text{ m}^2, 2.5\times 10^{-3} \cdot \mathbf{I}_4  \text{ rad}^2, \sigma_{\textsf{d}}^{2})$. We initialize the \ac{ue} at $[70.7285 \, \text{m},0 \, \text{m},\pi/2 \, \text{rad}, 300 \, \text{m}]^{\mathsf{T}}$ with covariance as $\text{diag}[ 0.3 \, \text{m}^{2},0.3 \, \text{m}^{2} \, 0.0052 \, \text{rad}^{2},0.3 \, \text{m}^{2}]$. The \ac{bs} is a priori known to the \ac{ue}. We implemented the EK-PMB SLAM filter with considering $\gamma=10$ best \acp{DA} every time step, and compared the results of cases using different levels of $\sigma_{\textsf{d}}$ and without considering Doppler. The mapping  performance is quantified by the \ac{gospa} distance \cite{rahmathullah2017generalized} for both \acp{va} and \acp{sp}, separately, the positioning performance is evaluated by the root mean squared error (RMSE), and we also compare bounds of different cases. More details on parametric settings of the filter can be found in \cite{ge2022computationally}.  The results are averaged over 100 Monte Carlo  simulations.

\subsection{Results and Discussion}
We firstly analyze how different levels of $\sigma_{\textsf{d}}$ affect the bounds. Fig.\,\ref{Fig.bounds} shows how the bounds 
change with $\sigma_{\textsf{d}}$, compared with the bounds without considering Doppler. We observe that involving Doppler as a dimension of the measurement improves the positioning and mapping performances of the \ac{slam} system, as the error bounds are lower than in the case where the Doppler measurement is ignored. This effect is stronger when Doppler measurements are more accurate, i.e., when $\sigma_{\textsf{d}}$ is smaller. 

\begin{figure}[t]
\center
\definecolor{mycolor1}{rgb}{0.00000,0.44700,0.74100}%
\definecolor{mycolor2}{rgb}{0.85000,0.32500,0.09800}%
\definecolor{mycolor3}{rgb}{0.92900,0.69400,0.12500}%
\definecolor{mycolor4}{rgb}{0.49400,0.18400,0.55600}%
\definecolor{mycolor5}{rgb}{0,0,0}%
%
\begin{tikzpicture}[scale=0.8\linewidth/14cm]

\begin{axis}[%
width=6.028in,
height=2.709in,
at={(1.011in,2.014in)},
scale only axis,
xmin=0.05,
xmax=0.5,
xlabel style={font=\color{white!15!black},font=\Large},
xlabel={$\sigma_{\textsf{d}}$ [m/s]},
ymin=0,
ymax=1,
legend columns=2,
ymode=log,
ylabel style={font=\color{white!15!black},font=\Large},
ylabel={PEB [m], HEB [deg], CEB [m], LEB [m]},
axis background/.style={fill=white},
axis x line*=bottom,
axis y line*=left,
legend style={legend cell align=left, align=left, draw=white!15!black,font=\Large}
]

\addplot [color=mycolor1,  line width=2.0pt]
  table[row sep=crcr]{%
0.05	0.157025987486338\\
0.1	0.171294347939749\\
0.2	0.18436581576857\\
0.3	0.189569309272586\\
0.4	0.192104817219046\\
0.5	0.192856513210559\\
};
\addlegendentry{PEB, with Doppler}

\addplot [color=mycolor2,  line width=2.0pt]
  table[row sep=crcr]{%
0.05	0.119872781845828\\
0.1	0.139226062464489\\
0.2	0.167296959580763\\
0.3	0.186877576665398\\
0.4	0.201318562377767\\
0.5	0.211101087992358\\
};
\addlegendentry{HEB, with Doppler}

\addplot [color=mycolor3,  line width=2.0pt]
  table[row sep=crcr]{%
0.05	0.0500062995470497\\
0.1	0.057868922464725\\
0.2	0.0668358826744384\\
0.3	0.0704536112828206\\
0.4	0.0720745693118308\\
0.5	0.0732899135005034\\
};
\addlegendentry{CEB, with Doppler}

\addplot [color=mycolor4,  line width=2.0pt]
  table[row sep=crcr]{%
0.05	0.430769231355449\\
0.1	0.583111762817011\\
0.2	0.732071384046513\\
0.3	0.787011243739541\\
0.4	0.815502184521895\\
0.5	0.830138663021057\\
};
\addlegendentry{LEB, with Doppler}

\addplot [color=mycolor1,dashed, line width=2.0pt]
  table[row sep=crcr]{%
0.05	0.193390889148602\\
0.1	0.193390889148602\\
0.2	0.193390889148602\\
0.3	0.193390889148602\\
0.4	0.193390889148602\\
0.5	0.193390889148602\\
};
\addlegendentry{PEB, without Doppler}

\addplot [color=mycolor2,dashed, line width=2.0pt]
  table[row sep=crcr]{%
0.05	0.243614338605646\\
0.1	0.243614338605646\\
0.2	0.243614338605646\\
0.3	0.243614338605646\\
0.4	0.243614338605646\\
0.5	0.243614338605646\\
};
\addlegendentry{HEB, without Doppler}

\addplot [color=mycolor3,dashed, line width=2.0pt]
  table[row sep=crcr]{%
0.05	0.0741269200513048\\
0.1	0.0741269200513048\\
0.2	0.0741269200513048\\
0.3	0.0741269200513048\\
0.4	0.0741269200513048\\
0.5	0.0741269200513048\\
};
\addlegendentry{CEB, without Doppler}

\addplot [color=mycolor4,dashed, line width=2.0pt]
  table[row sep=crcr]{%
0.05	0.846768182344354\\
0.1	0.846768182344354\\
0.2	0.846768182344354\\
0.3	0.846768182344354\\
0.4	0.846768182344354\\
0.5	0.846768182344354\\
};
\addlegendentry{LEB, without Doppler}

\end{axis}
\end{tikzpicture}%
\vspace{-5mm}
\caption{The \ac{PEB},  the \ac{HEB}, the \ac{CEB} and the average \ac{LEB} change with $\sigma_{\textsf{d}}$. The benchmarks are the bounds of the case without considering Doppler.}\vspace{-4mm}
\label{Fig.bounds}
\end{figure}
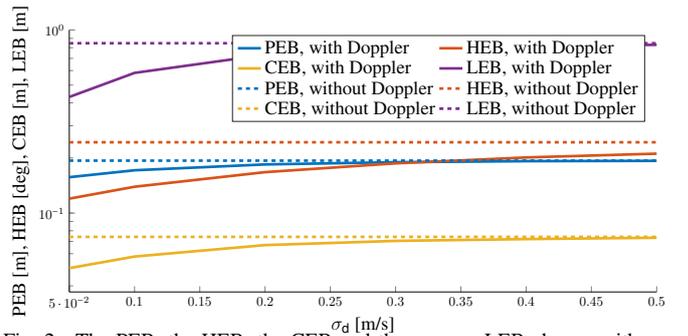

\begin{figure}[t]
\center
\definecolor{mycolor1}{rgb}{0.00000,0.44700,0.74100}%
\definecolor{mycolor2}{rgb}{0.85000,0.32500,0.09800}%
\definecolor{mycolor3}{rgb}{0.00000,0.44700,0.74100}%
\definecolor{mycolor4}{rgb}{0.85000,0.32500,0.09800}%
\definecolor{mycolor5}{rgb}{0,0,0}%
%
\begin{tikzpicture}[scale=0.8\linewidth/14cm]

\begin{axis}[%
width=6.028in,
height=2.009in,
at={(1.011in,2.014in)},
scale only axis,
xmin=0,
xmax=40,
xlabel style={font=\color{white!15!black},font=\Large},
xlabel={time step},
ymin=0,
ymax=30,
ylabel style={font=\color{white!15!black},font=\Large},
ylabel={GOSPA distance [m]},
axis background/.style={fill=white},
axis x line*=bottom,
axis y line*=left,
legend style={legend cell align=left, align=left, draw=white!15!black,font=\Large}
]

\addplot [color=mycolor1,  line width=2.0pt]
  table[row sep=crcr]{%
1	28.2842712474619\\
2	16.1613435774681\\
3	11.267198241046\\
4	8.50866436879003\\
5	7.42451698701259\\
6	7.08728659682638\\
7	6.78737174237077\\
8	6.59057937564529\\
9	6.24655802742255\\
10	5.38751428200149\\
11	4.98160911941193\\
12	4.56452382949523\\
13	4.46400551204806\\
14	4.22392419322941\\
15	3.98353198650708\\
16	3.91947304750308\\
17	3.91154838663654\\
18	3.89494749581814\\
19	3.90745345678338\\
20	3.85787218431992\\
21	3.81833010155553\\
22	3.74345371911451\\
23	3.67296859093222\\
24	3.64221002840471\\
25	3.55146691443283\\
26	3.5720481374798\\
27	3.56489387914315\\
28	3.98073144369057\\
29	4.00924818474199\\
30	3.65970809496629\\
31	3.7172814604156\\
32	3.74130063535291\\
33	3.71912092165811\\
34	3.71763780926782\\
35	3.71987787713517\\
36	3.71313595115809\\
37	3.69566176267095\\
38	3.66690683931629\\
39	3.65239121498151\\
40	3.65191304908262\\
};
\addlegendentry{With Doppler}


\addplot [color=mycolor1,dashed, line width=2.0pt]
  table[row sep=crcr]{%
1	28.2842712474619\\
2	14.8574127885984\\
3	12.7114643969106\\
4	10.9455764472861\\
5	9.09012368315244\\
6	8.19212760858123\\
7	7.54957976744773\\
8	9.28439690903019\\
9	9.16834610469031\\
10	9.1099994789594\\
11	8.8894910172174\\
12	7.86868730646592\\
13	7.32042763833874\\
14	6.64546221142368\\
15	6.57706623431597\\
16	6.45400658595155\\
17	6.28937601677533\\
18	8.50577551841823\\
19	9.18212437479686\\
20	9.29004549810707\\
21	9.30899544410811\\
22	8.67595945185318\\
23	7.14562082330056\\
24	6.92017350505839\\
25	6.98461825617181\\
26	7.00164041350459\\
27	7.05715961035402\\
28	7.82090188559817\\
29	8.85820434756982\\
30	8.91007313184357\\
31	8.36193659399435\\
32	7.36969202153649\\
33	7.38049919567709\\
34	7.23372870201308\\
35	7.21042169236854\\
36	7.16209264617936\\
37	7.13458206030356\\
38	7.12977893574705\\
39	7.27975245572636\\
40	7.65643732957054\\
};
\addlegendentry{Without Doppler}


\end{axis}
\end{tikzpicture}%
\vspace{-5mm}
\caption{Comparison of mapping performances for VAs between two cases: with and without considering Doppler.}\vspace{-5mm}
\label{Fig.mapping}
\end{figure}
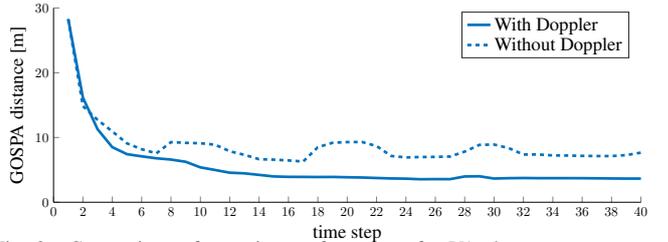

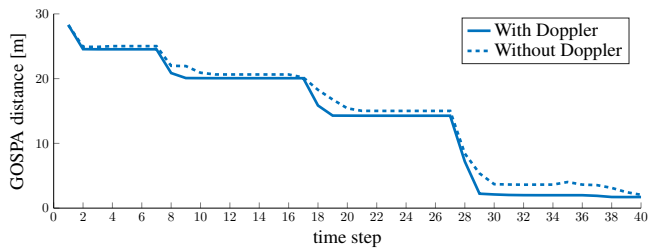
\begin{figure}
\center
\definecolor{mycolor1}{rgb}{0.00000,0.44700,0.74100}%
\definecolor{mycolor2}{rgb}{0.85000,0.32500,0.09800}%
\definecolor{mycolor3}{rgb}{0.00000,0.44700,0.74100}%
\definecolor{mycolor4}{rgb}{0.85000,0.32500,0.09800}%
\definecolor{mycolor5}{rgb}{0,0,0}%
%
\begin{tikzpicture}[scale=0.8\linewidth/14cm]

\begin{axis}[%
width=6.028in,
height=2.009in,
at={(1.011in,2.014in)},
scale only axis,
xmin=0,
xmax=40,
xlabel style={font=\color{white!15!black},font=\Large},
xlabel={time step},
ymin=0,
ymax=30,
ylabel style={font=\color{white!15!black},font=\Large},
ylabel={GOSPA distance [m]},
axis background/.style={fill=white},
axis x line*=bottom,
axis y line*=left,
legend style={legend cell align=left, align=left, draw=white!15!black,font=\Large}
]

\addplot [color=mycolor1,  line width=2.0pt]
  table[row sep=crcr]{%
1	28.2842712474619\\
2	24.5501266787321\\
3	24.5300823835694\\
4	24.5300823835694\\
5	24.5300823835694\\
6	24.5300823835694\\
7	24.5300823835694\\
8	20.8602750449368\\
9	20.0868704160714\\
10	20.0729241617819\\
11	20.0655570449193\\
12	20.0649300164732\\
13	20.0645193870004\\
14	20.0645193870004\\
15	20.0645193870004\\
16	20.0645193870004\\
17	20.0645193870004\\
18	15.8494445707689\\
19	14.2894045965689\\
20	14.2821805318767\\
21	14.2742673418575\\
22	14.2724741994292\\
23	14.2700655568798\\
24	14.2700655568798\\
25	14.2700655568798\\
26	14.2700655568798\\
27	14.2700655568798\\
28	7.19754415511525\\
29	2.22960668122651\\
30	2.11279955103129\\
31	2.03265580068579\\
32	2.00342818568324\\
33	1.99561494086911\\
34	1.99561494086911\\
35	1.99561494086911\\
36	1.99561494086911\\
37	1.90151446111805\\
38	1.71930573523988\\
39	1.70809523148783\\
40	1.71478414532627\\
};
\addlegendentry{With Doppler}

\addplot [color=mycolor1,dashed, line width=2.0pt]
  table[row sep=crcr]{%
1	28.2842712474619\\
2	24.9131056182981\\
3	24.9057497288635\\
4	25.0170001471354\\
5	25.0170001471354\\
6	25.0170001471354\\
7	25.0170001471354\\
8	21.9812277919597\\
9	21.938248604687\\
10	20.9084316897639\\
11	20.64402222741\\
12	20.6416151735271\\
13	20.6406883984759\\
14	20.6406886316049\\
15	20.6406886316049\\
16	20.6406886316049\\
17	20.1757951086278\\
18	18.2658345128125\\
19	16.7721269062796\\
20	15.4121191399715\\
21	15.0214321632327\\
22	15.0158069936847\\
23	15.0135859222393\\
24	15.0136111474607\\
25	15.013795524628\\
26	15.0138255708017\\
27	15.0138255708017\\
28	8.42772235752995\\
29	5.3675251942532\\
30	3.70120321380309\\
31	3.64615628697128\\
32	3.62032819777826\\
33	3.63255541941277\\
34	3.63255541941277\\
35	4.03156676520894\\
36	3.63255541941277\\
37	3.55186422503854\\
38	3.12436465365146\\
39	2.4681067662366\\
40	2.05300503755635\\
};
\addlegendentry{Without Doppler}

\end{axis}
\end{tikzpicture}%
\caption{Comparison of mapping performances for SPs between two cases: with and without considering Doppler.}
\label{Fig.mapping_SP}
\end{figure}

\begin{figure}
\center
%
%
\definecolor{mycolor1}{rgb}{0.00000,0.44700,0.74100}%
\definecolor{mycolor2}{rgb}{0.85000,0.32500,0.09800}%
\definecolor{mycolor3}{rgb}{0.92900,0.69400,0.12500}%
\definecolor{mycolor4}{rgb}{0.49400,0.18400,0.55600}
\definecolor{mycolor5}{rgb}{0,0,0}%
\begin{tikzpicture}[scale=0.99\linewidth/14cm]

\begin{axis}[%
width=3.842in,
height=1.281in,
at={(3.465in,2.378in)},
scale only axis,
bar shift auto,
xmin=0.5,
xmax=3.5,
xtick={1,2,3},
xticklabels={{position},{heading},{clock bias}},
ymin=0,
ymax=0.7,
ylabel style={font=\color{white!15!black}},
ylabel={RMSE of state [m], [deg], [m]},
axis background/.style={fill=white},
legend style={ anchor=north east, legend cell align=left, align=left, draw=white!15!black}
]

\addplot[ybar, bar width=0.145, fill=mycolor1, draw=black, area legend] table[row sep=crcr] {%
1	0.5314\\
2	0.3560\\
3	0.1849\\
};
\addplot[forget plot, color=white!15!black] table[row sep=crcr] {%
0.5	0\\
3.5	0\\
};
\addlegendentry{With Doppler}

\addplot[ybar, bar width=0.145, fill=mycolor2, draw=black, area legend] table[row sep=crcr] {%
1	0.6194\\
2	0.4021\\
3	0.2458\\
};
\addplot[forget plot, color=white!15!black] table[row sep=crcr] {%
0.5	0\\
3.5	0\\
};
\addlegendentry{Without Doppler}

\end{axis}
\end{tikzpicture}%
\caption{Comparison of \ac{ue} state estimation between two cases: with and without considering Doppler.}
\label{Fig.bar}
\end{figure}
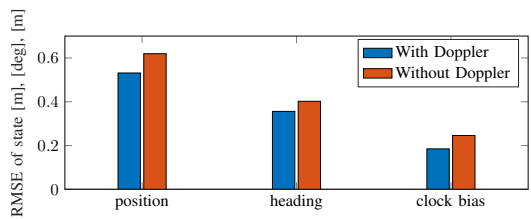

To validate the benefits of Doppler in mmWave radio SLAM,  we then implement the EK-PMB SLAM filter on two cases: 1) involving Doppler into the measurement with $\sigma_{\textsf{d}}=0.1 \text{ m/s}$ and 2) without considering Doppler, and compare the mapping performance for both \acp{va} and \acp{sp}, and the positioning performance between two cases, as illustrated in Fig.\,\ref{Fig.mapping}, Fig.\,\ref{Fig.mapping_SP} and Fig.\,\ref{Fig.bar}, respectively. Fig.\,\ref{Fig.mapping} and Fig.\,\ref{Fig.mapping_SP} demonstrate that the SLAM filter can map the environment for both cases, and the mapping accuracy improves with more measurements being received, as overall, all GOSPA distances gradually decrease over time.  Clearly, when Doppler is involved, the SLAM filter has better mapping performance, as the solid lines are lower than the dashed lines in both Fig.\,\ref{Fig.mapping} and Fig.\,\ref{Fig.mapping_SP}. Fig.\,\ref{Fig.bar} indicates that considering Doppler results in better \ac{ue} state estimates, as lower RMSEs can be acquired. The reasons are that considering Doppler as a dimension of the measurement brings lower bounds. In addition, it also helps the SLAM filter to solve the \ac{DA} problem, which improves the average weight for the correct \ac{DA} from 0.6763 to 0.8622, with weights of all selected $\gamma=10$ \acp{DA} summed to 1 every time step.

\section{Conclusions}
In this paper, we exploited Doppler as a part of measurement in bistatic mmWave radio SLAM, formulated the EK-PMB SLAM filter on the new measurement model, and provided the \ac{PCRB} for the model. Our results theoretically indicate that the involvement of Doppler helps the SLAM filter to acquire better mapping and positioning performance than the case without considering Doppler. The better Doppler observations are, the lower \acp{PCRB} and the more accurate  map and \ac{ue} state estimates will be. The implementation of the EK-PMB SLAM filter validates the theoretical benefits of involving  Doppler. Future work will include extending the \ac{ue} model to a model with unknown speed, the inclusion of high-dimensional channel estimation, as well as extending the \ac{slam} problem to a \ac{slat} problem.

\scriptsize{
\section*{Acknowledgment}
This work was partially supported by the Wallenberg AI, Autonomous Systems and Software Program (WASP) funded by Knut and Alice Wallenberg Foundation, and the Vinnova 5GPOS project under grant 2019-03085, by the Swedish Research Council under grant 2018-03705, and by the Academy of Finland under the grants \#315858, \#328214, \#319994, \#323244, \#346622.
}
\balance

\bibliography{IEEEabrv,Bibliography}

\begin{thebibliography}{10}
\providecommand{\url}[1]{#1}
\csname url@rmstyle\endcsname
\providecommand{\newblock}{\relax}
\providecommand{\bibinfo}[2]{#2}
\providecommand\BIBentrySTDinterwordspacing{\spaceskip=0pt\relax}
\providecommand\BIBentryALTinterwordstretchfactor{4}
\providecommand\BIBentryALTinterwordspacing{\spaceskip=\fontdimen2\font plus
\BIBentryALTinterwordstretchfactor\fontdimen3\font minus
  \fontdimen4\font\relax}
\providecommand\BIBforeignlanguage[2]{{%
\expandafter\ifx\csname l@#1\endcsname\relax
\typeout{** WARNING: IEEEtran.bst: No hyphenation pattern has been}%
\typeout{** loaded for the language `#1'. Using the pattern for}%
\typeout{** the default language instead.}%
\else
\language=\csname l@#1\endcsname
\fi
#2}}
\renewcommand\BIBentryALTinterwordstretchfactor{4}

\bibitem{nurmi2017multi}
J.~Nurmi, E.-S. Lohan, H.~Wymeersch, G.~Seco-Granados, and O.~Nyk{\"a}nen,
  \emph{Multi-Technology Positioning}.\hskip 1em plus 0.5em minus 0.4em\relax
  Springer, 2017.

\bibitem{patwari2005}
N.~Patwari, J.~Ash, S.~Kyperountas, A.~Hero, R.~Moses, and N.~Correal,
  ``Locating the nodes: cooperative localization in wireless sensor networks,''
  \emph{IEEE Signal Processing Magazine}, vol.~22, no.~4, pp. 54--69, 2005.

\bibitem{larsen2009}
M.~D. Larsen, A.~L. Swindlehurst, and T.~Svantesson, ``Performance bounds for
  {MIMO-OFDM} channel estimation,'' \emph{IEEE Transactions on Signal
  Processing}, vol.~57, no.~5, pp. 1901--1916, 2009.

\bibitem{wymeersch20175g}
H.~Wymeersch, G.~Seco-Granados, G.~Destino, D.~Dardari, and F.~Tufvesson,
  ``{5G} {mmWave} positioning for vehicular networks,'' \emph{IEEE Wireless
  Communications}, vol.~24, no.~6, pp. 80--86, 2017.

\bibitem{patole2017automotive}
S.~M. Patole, M.~Torlak, D.~Wang, and M.~Ali, ``Automotive radars: {A} review
  of signal processing techniques,'' \emph{IEEE Signal Processing Magazine},
  vol.~34, no.~2, pp. 22--35, 2017.

\bibitem{amar2008localization}
A.~Amar and A.~J. Weiss, ``Localization of narrowband radio emitters based on
  doppler frequency shifts,'' \emph{IEEE Transactions on Signal Processing},
  vol.~56, no.~11, pp. 5500--5508, 2008.

\bibitem{Han2016}
Y.~Han, Y.~Shen, X.-P. Zhang, M.~Z. Win, and H.~Meng, ``Performance limits and
  geometric properties of array localization,'' \emph{IEEE Transactions on
  Information Theory}, vol.~62, no.~2, pp. 1054--1075, 2016.

\bibitem{kakkavas2019performance}
A.~Kakkavas, M.~H.~C. Garc{\'\i}a, R.~A. Stirling-Gallacher, and J.~A. Nossek,
  ``Performance limits of single-anchor millimeter-wave positioning,''
  \emph{IEEE Trans. Wireless Commun.}, vol.~18, no.~11, pp. 5196--5210, Aug.
  2019.

\bibitem{qian2017widar}
K.~Qian, C.~Wu, Z.~Yang, Y.~Liu, and K.~Jamieson, ``Widar: {Decimeter}-level
  passive tracking via velocity monitoring with commodity {Wi-Fi},'' in
  \emph{Proceedings of the 18th ACM International Symposium on Mobile Ad Hoc
  Networking and Computing}, 2017.

\bibitem{wen20195g}
F.~Wen, J.~Kulmer, K.~Witrisal, and H.~Wymeersch, ``{5G} positioning and
  mapping with diffuse multipath,'' \emph{IEEE Transactions on Wireless
  Communications}, vol.~20, no.~2, pp. 1164--1174, 2020.

\bibitem{yassin2018mosaic}
A.~Yassin, Y.~Nasser, A.~Y. Al-Dubai, and M.~Awad, ``{MOSAIC}: {Simultaneous}
  localization and environment mapping using mmwave without a-priori
  knowledge,'' \emph{IEEE Access}, vol.~6, pp. 68\,932--68\,947, 2018.

\bibitem{Erik_BPSLAM_TWC2019}
E.~Leitinger, F.~Meyer, F.~Hlawatsch, K.~Witrisal, F.~Tufvesson, and M.~Z. Win,
  ``A belief propagation algorithm for multipath-based {SLAM},'' \emph{IEEE
  Trans. Wireless Commun.}, vol.~18, no.~12, pp. 5613--5629, Sep. 2019.

\bibitem{Rico_BPSLAM_JSTSP2019}
R.~Mendrzik, F.~Meyer, G.~Bauch, and M.~Z. Win, ``{Enabling situational
  awareness in millimeter wave massive MIMO systems},'' \emph{IEEE J. Sel.
  Topics Signal Process.}, vol.~13, no.~5, pp. 1196--1211, Aug. 2019.

\bibitem{kim20205g}
H.~Kim, K.~Granstr{\"o}m, L.~Gao, G.~Battistelli, S.~Kim, and H.~Wymeersch,
  ``{5G} {mmWave} cooperative positioning and mapping using multi-model {PHD}
  filter and map fusion,'' \emph{IEEE Transactions on Wireless Communications},
  2020.

\bibitem{ge20205GSLAM}
\BIBentryALTinterwordspacing
Y.~Ge, F.~Wen, H.~Kim, M.~Zhu, F.~Jiang, S.~Kim, L.~Svensson, and H.~Wymeersch,
  ``{5G} {SLAM} using the clustering and assignment approach with diffuse
  multipath,'' \emph{Sensors (Basel, Switzerland)}, vol.~20, no.~16, August
  2020. [Online]. Available: \url{https://doi.org/10.3390/s20164656}
\BIBentrySTDinterwordspacing

\bibitem{ge2022computationally}
Y.~Ge, O.~Kaltiokallio, H.~Kim, F.~Jiang, J.~Talvitie, M.~Valkama, L.~Svensson,
  S.~Kim, and H.~Wymeersch, ``A computationally efficient {EK-PMBM} filter for
  bistatic {mmWave} radio {SLAM},'' \emph{IEEE Journal on Selected Areas in
  Communications}, 2022.

\bibitem{roth2014}
M.~{Roth}, G.~{Hendeby}, and F.~{Gustafsson}, ``{EKF/UKF} maneuvering target
  tracking using coordinated turn models with polar/cartesian velocity,'' in
  \emph{17th International Conference on Information Fusion}, 2014, pp. 1--8.

\bibitem{heath2016overview}
R.~W. Heath, N.~Gonzalez-Prelcic, S.~Rangan, W.~Roh, and A.~M. Sayeed, ``An
  overview of signal processing techniques for millimeter wave {MIMO}
  systems,'' \emph{IEEE Journal of Selected Topics in Signal Processing},
  vol.~10, no.~3, pp. 436--453, 2016.

\bibitem{richter2005estimation}
A.~Richter, ``Estimation of radio channel parameters: Models and algorithms,''
  Ph.D. dissertation, Ilmenau University of Technology, 2005.

\bibitem{alkhateeb2014channel}
A.~Alkhateeb, O.~El~Ayach, G.~Leus, and R.~W. Heath, ``Channel estimation and
  hybrid precoding for millimeter wave cellular systems,'' \emph{IEEE Journal
  of Selected Topics in Signal Processing}, vol.~8, no.~5, pp. 831--846, 2014.

\bibitem{jiang2021high}
F.~Jiang, Y.~Ge, M.~Zhu, and H.~Wymeersch, ``High-dimensional channel
  estimation for simultaneous localization and communications,'' in \emph{IEEE
  Wireless Communications and Networking Conference (WCNC)}, 2021.

\bibitem{van2004detection}
H.~L. Van~Trees, \emph{Detection, Estimation, and Dodulation Theory, Part I:
  Detection, Estimation, and Linear Modulation Theory}.\hskip 1em plus 0.5em
  minus 0.4em\relax John Wiley \& Sons, 2004.

\bibitem{tichavsky1998}
P.~{Tichavsky}, C.~H. {Muravchik}, and A.~{Nehorai}, ``Posterior
  {C}ram{\'e}r-{R}ao bounds for discrete-time nonlinear filtering,'' \emph{IEEE
  Transactions on Signal Processing}, vol.~46, no.~5, pp. 1386--1396, 1998.

\bibitem{EKPHD2021Ossi}
O.~Kaltiokallio, Y.~Ge, J.~Talvitie, H.~Wymeersch, and M.~Valkama, ``{mmWave}
  simultaneous localization and mapping using a computationally efficient
  {EK-PHD} filter,'' in \emph{IEEE International Conference on Information
  Fusion (Fusion)}, 2021.

\bibitem{garcia2018poisson}
{\'A}.~F. Garc{\'\i}a-Fern{\'a}ndez, J.~L. Williams, K.~Granstr{\"o}m, and
  L.~Svensson, ``Poisson {multi-Bernoulli} mixture filter: {Direct} derivation
  and implementation,'' \emph{IEEE Transactions on Aerospace and Electronic
  Systems}, vol.~54, no.~4, pp. 1883--1901, 2018.

\bibitem{williams2015marginal}
J.~L. Williams, ``Marginal multi-{Bernoulli} filters: {RFS} derivation of
  {MHT}, {JIPDA}, and association-based {MeMBer},'' \emph{IEEE Transactions on
  Aerospace and Electronic Systems}, vol.~51, no.~3, pp. 1664--1687, 2015.

\bibitem{rahmathullah2017generalized}
A.~S. Rahmathullah, {\'A}.~F. Garc{\'\i}a-Fern{\'a}ndez, and L.~Svensson,
  ``Generalized optimal sub-pattern assignment metric,'' in \emph{20th IEEE
  International Conference on Information Fusion (Fusion)}, 2017.

\end{thebibliography}



%

\end{document}